\documentclass[11pt,a4paper]{article}
\usepackage{amsmath}
\usepackage{dsfont}
\usepackage{bm}
\usepackage{esint}
\usepackage{subcaption}
\usepackage{accents}
\usepackage{mathrsfs}
\usepackage{amsbsy}

\usepackage{a4wide,graphicx,times,psfrag,wrapfig,sidecap}
\usepackage{cite}
\usepackage[colorlinks=true,linkcolor=black, citecolor=black,
urlcolor=black]{hyperref}
\numberwithin{equation}{section}
\makeatletter \let\old@startsection=\@startsection
\renewcommand{\@startsection}[6]
{\old@startsection{#1}{#2}{#3}{#4}{#5}{#6\mathversion{bold}}}
\makeatother

\def\<{\langle}
\def\>{\rangle}

\def\tr{{\rm   tr} }
\def\Im{{\rm Im}}
\def\Re{{\rm Re}}
\newcommand\encadremath[1]{\vbox{\hrule\hbox{\vrule\kern8pt
\vbox{\kern8pt \hbox{$\displaystyle #1$}\kern8pt}
\kern8pt\vrule}\hrule}} \def\enca#1{\vbox{\hrule\hbox{
\vrule\kern8pt\vbox{\kern8pt \hbox{$\displaystyle #1$} \kern8pt}
\kern8pt\vrule}\hrule}}
  \usepackage{bm}

\def\XXint#1#2#3{{\setbox0=\hbox{$#1{#2#3}{\int}$}
     \vcenter{\hbox{$#2#3$}}\kern-.5\wd0}}

\newcommand\antifrac[2]{\frac{#2}{#1}}

\begin{document}

\begin{center}
{\Large  Logarithmic Negativity  and Spectrum in Free Fermionic Systems for Well-separated Intervals }

\vspace{10mm}

Eldad Bettelheim  \\[7mm]

Racah Institute of Physics, Hebrew University of Jerusalem, Edmund J Safta Campus\\ 91904 Jerusalem, Israel\\ 

\vspace{20mm}

\end{center}

\vskip9mm

\noindent{ We employ a mathematical framework based on the Riemann-Hilbert approach developed in Ref. \cite{Bettelheim:Banerjee:Plenio} to  study  logarithmic negativity of two intervals of free fermions  in the case where  the size of the intervals as well as the distance between them is macroscopic.  We find that  none of the eigenvalues of the density matrix become negative, but rather they develop a small imaginary value, leading to non-zero logarithmic negativity. As an example, we compute negativity at half-filling and for intervals of equal size we find a result of order $(\log(N))^{-1}$, where $N$\ is the typical length scale in units of the lattice spacing. One may compute logarithmic negativity in further situations, but we find that the results are non-universal, depending non-smoothly on the Fermi level and the size of the intervals in  units of the lattice spacing.         }

\section{Introduction and Results}
The object of the this paper is to study  logarithmic negativity\cite{Eisert:Cramer:Plenio:Colloquium,plenio:Logarithmic:Negativity:Monotone,eisert-plenio:Comparison} of two intervals of  free fermions for non-adjacent intervals. This is to be contrasted to previous calculations where explicit results are available for fermions when the intervals are adjacent\cite{Calabrese:Cardy:Tonni:Negativity:Field:Theoretic,Calabrese:Cardy:Tonni:Negativity:QFT,Calabrese:Cardy:Tonni:Negativity:Two:Intervals:I,Calabrese:Cardy:Tonni:Negativity:Two:Intervals:II}. In a previous publication\cite{Bettelheim:Banerjee:Plenio} we have shown how to map this question into a Riemann-Hilbert problem using the orthogonal polynomial technique following Refrs. \cite{Deift:Its:Krasovksy:Toeplitz:Hankel,Its:Krasovsky:Gaussian:With:Kump,Deift:Its:Krasovsky:Toeplitz} which use this method to prove the Fisher-Hartwig theorem\cite{fisher:hartwig:Theorem}, the latter being intimately related to entanglement measures of free translationally invariant systems\cite{Jin:Korepin:Entanglement:Entropy:Fermions}.  We show here that the associated Riemann-Hilbert problem is solvable in the limit of macroscopic intervals (namely, the thermodynamic $N\to\infty$ limit where $N$ is the typical number of sites in the intervals or between the intervals) and that this technique may be used to compute the location of the individual eigenvalues of the quantum density matrix following Ref. \cite{Deift:Its:Krasovsky}. 

After displaying the solution of the Riemann-Hilbert problem in this case, we proceed to compute the negativity spectrum and logarithmic negativity. In order to do so, we must compute the fine structure of the eigenvalues of the deformed correlation matrix (where one of the intervals is time reversed). In ``fine structure" we mean that the exact position of the eigenvalues must be found. This is required since we must compute the small imaginary part of the eigenvalues which in fact fully holds the information about  negativity in this case, as we find no eigenvalues which are actually negative. To find this fine structure, we employ slightly more subtle Riemann-Hilbert methods combined with orthogonal polynomial identities, which we develop for this case. 

The final result shows that negativity displays non-universal behavior. In particular, negativity for the case of a zero-temperature state, depends in a non-smooth way on the Fermi momentum of the fermions, as well as on the length of the intervals, featuring a on the relevant lengths which may have, for example, even-odd effects. For example at half filling (the Fermi sea occupies half of the Brillouin zone), and with equal lengths of the intervals, negativity depends on the parity of the distance between the intervals. We find non-zero negativity only in the case where this distance holds an odd number of sites. Denoting the endpoints of intervals as $-t_i$  where $i$ runs from $0$ to $3$ from  right to right (see Fig. \ref{System}), negativity for half filling and equal size of the intervals and odd number of sites between the intervals and between the two outer most points takes the simple form:
\begin{align}
\label{FinalFinal}\mathcal{E}=\frac{1}{4}\frac{\log^2\left[\frac{\ell_{30}}{\ell_{21}}\right]}{\log\antifrac{\ell_{30}\ell_{21}}{4\ell^2_{20}\ell^2_{10}}}.
\end{align}
Where $\ell_{ij}=|t_i-t_j|$ is the distance between point $i$ and $j,$  The result shows that in this limit negativity, denoted by $\mathcal{E}$ behaves as $\frac{1}{\log(N)}$.

Using the formulas developed below,  negativity may be computed in all other cases to any given order for as long as all relevant distances are thermodynamic ($N\to\infty$) and the Fermi momentum is finite, however due to the non-universal nature of the result we restrict ourselves to specific cases where the final result is simple and transparent as in Eq. (\ref{FinalFinal}) or is simply zero.

\section{Entanglement Entropy and Negativity}
We shall deal exclusively with translationally invariant fermionic systems on the infinite line, such that we have a covariance function $f_k$ defined below:
\begin{align}
f_{i-j}=\tr \left((2\hat\rho-\mathds{1}) c^\dagger_ic_j\right),
\end{align}
along with its Fourier transform $f(z)$ defined as follows:
\begin{align}
f(z)=\sum_i f_i z^i.
\end{align}
$f(z)$ is actually the Fermi function, namely the occupation number  of the fermions in momentum space as is given below: \begin{align}
\<c^\dagger(p) c(p)\>\equiv\tr \left(\hat\rho c^\dagger(p)c(p)\right)=\frac{f(e^{\imath p})+1}{2}.\label{FermiFunction}
\end{align}
A zero temperature Fermi surface is characterized by a jump discontinuity of $f(p)$ at the Fermi points.

Momenta are measured here in the Brillouin zone. Namely, $p\in[-\pi,\pi]$. Zero temperature states are characterized by a Fermi momentum $p_F\in[-\pi,\pi]$, which characterizes a Fermi sea with two Fermi points, which are  $p_F$ and  $-p_F$. The filling fraction is defined by $\frac{p_F}{\pi}$.    

If we have a two intervals $A$ and $B$ one may consider the reduced density matrix $\hat{\rho}_{A\cup B}=\tr_{(A\cup B)^C}(\hat \rho)$, where the trace is taken only over the complement of $A\cup B$. In this case the density function $f$ trivially remains the same when it is restricted to  the intervals $A\cup B$ however the reduced density matrix $\hat{\rho}_{A\cup B}$ has different eigenvalues due to the restriction. The  matrix $\hat{\rho}_{A\cup B}$ remains Gaussian since the correlations continue to obey Wick's theorem\cite{Peschel:Correlation} (since they are unchanged). It is then easy to see that there is a relation between the eigenvalues of the matrix whose $(i,j)$ element is  $f_{i-j}$ where $i$ and $j$ are restricted to the intervals $A\cup B$\ and the eigenvalues of the matrix $\hat{\rho}_{A \cup B}$.      

In fact,  if there are $N$\ eigenvalues of the covariance matrix one finds $2^N$ eigenvalues of the density matrix. Specifically, let $N$\ be the total size of $A\cup B$, then if $\lambda_i$ are the $N$ eigenvalues of the $N\times N$ covariance matrix, then for any choice of $N$ values  for $\nu_i$, where $\nu_i\in\{\frac{1+\lambda_i}{2},\frac{1-\lambda_i}{2}\}$ (corresponding to the choice of occupied and unoccupied fermionic states, respectively),   there will correspond an eigenvalue of the density matrix, $\hat \rho_{A\cup B},$  of the form $
\prod_{i} \nu_i.
$
Despite of the fact that the eigenvalues of the density matrix are the \emph{ products }of the $\nu_i$, we  shall refer, in a somewhat loose manner, to the \emph{individual} values of $\nu_i,$ as the `eigenvalues of the density matrix' and denote them, as we have just done, by $\nu_i$. The spectrum of $\nu_i$ is known as the entanglement spectrum, and is related to the spectrum of $\lambda_i$ as follows 
$
\pm\lambda_i=2\nu_i-1.
$
 The $\pm$ sign allows one to obtain both eigenvalues of the density matrix,  $\nu_i$ and $1-\nu_i,$ from the same value of $\lambda_i$.

 \subsection{Deformation of the Reduced Density and Covariance Matrix}
As mentioned above, computing negativity involves applying a sort of time reverse transformation on one of the intervals. We follow Refs \cite{shapourian:Hasan:Shiozako:ken:Ryu:Shinsei:Negativity,Shapourian:Ruggiero:Ryu:Twisted:Untwiste:Negativity}  and apply this transformation by deforming the  covariance matrix $f_{i-j}$ into the matrix  $\tilde{f}_{k_i,k_j}$  as follows:

\begin{align}
\tilde f_{i,j}=\tr \left((\mathds{1}-2\rho )c^\dagger_i c_j\right)\times\left\{\begin{array}{lr}
1 & i\in A, j\in A\\
-1 & i\in B, j\in B\\
i & i\in A, j\in B\\
i & i\in B, j\in A\\ 
\end{array}\right.\label{deformingf} .
\end{align}
This matrix has an obvious block diagonal structure. Due to this, we choose to define a matrix as follows: \begin{align}
 {\bm{f}}(z)=\begin{pmatrix}f(z) & \imath f(z)z^n \\
\imath f(z)z^{-n} & - f(z)\\
\end{pmatrix} ,
\end{align}
where we have dropped the tilde on the left hand side, as we shall always refer to the deformed correlation matrix, and so the boldface suffices to distinguish this matrix. \

\begin{figure}[h!!!]
\begin{center}
\includegraphics[width=10cm]{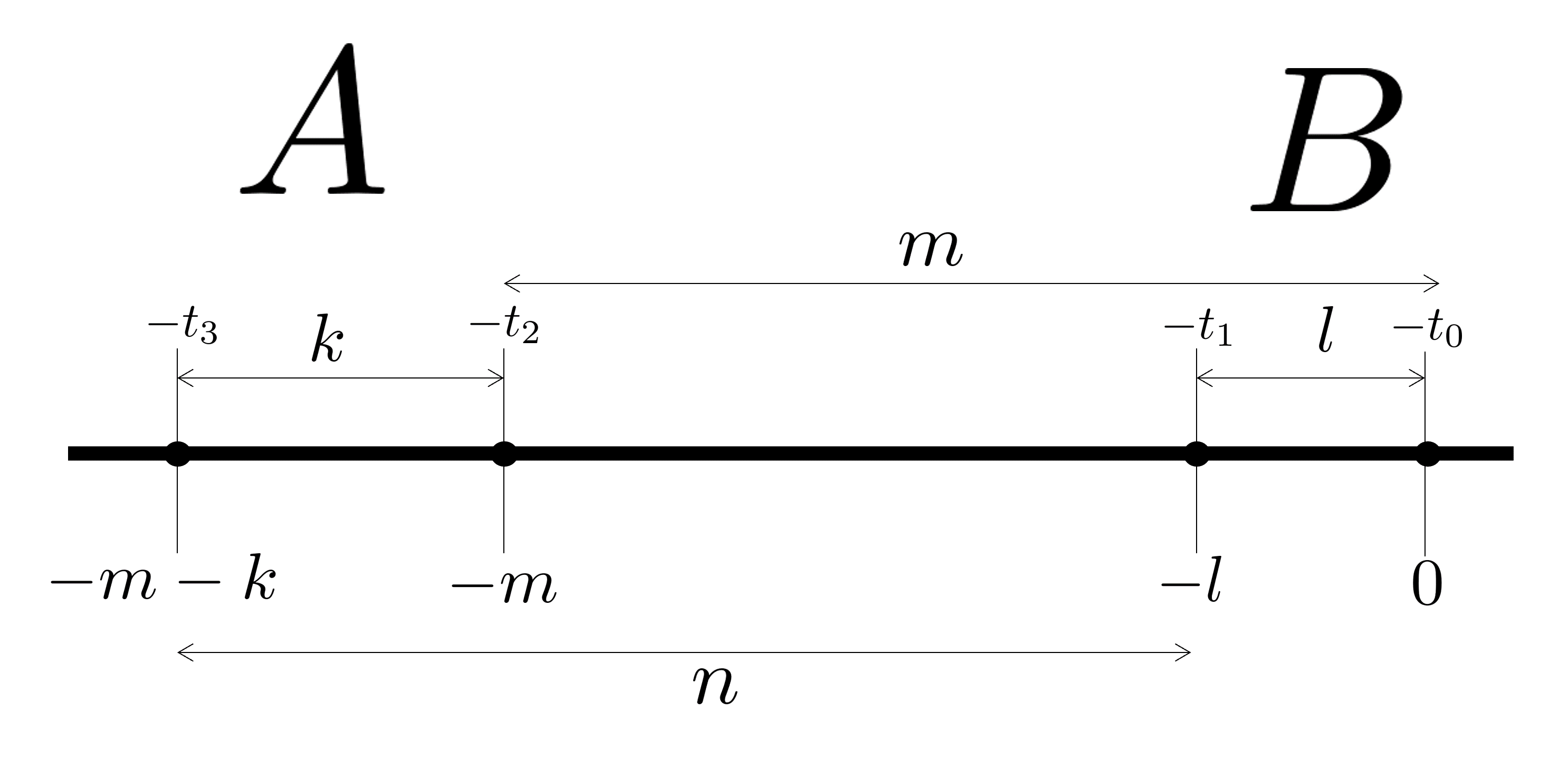}
\caption{
Two subsystems $A$ and $B$\ are identified on the 1 dimensional  real line (heavy line). The size of $A$ is $k+1$\ while the size of $B$\ is $l+1$ a. Further distances are denoted on the figure, such as $m$ and $n$ being the distance between the rightmost or leftmost points of the respective intervals, respectively. In addition, the points $t_i$ denote respective endpoints of the intervals.    \label{System}} \end{center}
\end{figure}

The sizes and different endpoints of the intervals are denoted in Fig. 
\ref{System} and shall be referred to in the following. 

The matrix $\bm f$ acts on a vector $\bm{\psi}=\begin{pmatrix} \psi_1(z) \\
\psi_2(z) \\
\end{pmatrix}$ by straightforward matrix multiplication. This encodes the matrix action of the whose elements matrix $\tilde{f}_{i,j}$ on a vector $\chi_j$ where $\chi_j $ is supported on $A\cup B$ (That is $\chi_j\neq0$ only if $j\in A\cup B$), if we choose $\bm \psi$ to be a vector of the following polynomials: 
\begin{align}
\psi_1(z)=\sum_{j\in A}\chi_j z^{j+t_3},\quad \psi_2(z)=\sum_{j\in B}\chi_j z^{j+t_1}, 
\end{align} 
such that the encoding is given by
\begin{align}
\mathcal{P}_{A\cup B}\bm f \bm \psi=\begin{pmatrix}\sum_{i\in A,j}z^{i+t_3} \tilde{f}_{i,j}\chi_j\\
\sum_{i\in B,j}z^{i+t_1} \tilde{f}_{i,j}\chi_j \\
\end{pmatrix},
\end{align}
where $\mathcal{P}_{A\cup B}$ is a projection operator ($\mathcal{P}^2_{A\cup B} =\mathcal{P}_{A\cup B}$) onto $A \cup B$. Namely,  $\mathcal{P}_{A\cup B}$ satisfies
\begin{align}
\quad \mathcal{P}_{A\cup B}\begin{pmatrix}z^i \\
0 \\
\end{pmatrix} = 0\quad  \mbox{ for }i+t_3\notin A ,\mbox{ and }\mathcal{P}_{A\cup B}\begin{pmatrix}0 \\
z^i \\
\end{pmatrix} = 0\quad  \mbox{ for }i+t_1\notin B. 
\end{align}We look for eigenvalues of the matrix $\bm f$:
\begin{align}
({\bm{f}}+ \lambda_i\mathds{1}) \bm\psi^{(-\lambda_i)}=0,\label{evEq}
\end{align} 
Once the roots of Eq. (\ref{evEq})\ are the eigenvalues of the deformed and reduced density matrix, which we denote  by $\nu_i.$ The relation between the eigenvalues of the correlation matrix and the density matrix  read as follows\cite{Peschel:Correlation}: 
\begin{align}
\pm \lambda_i=2\nu_i-1.\label{nulambdanega}
\end{align} 
This relation is a basic consequence of the fermionic algebra. 

Knowledge of the spectrum, $\lambda_i$ completely determines  logarithmic negativity defined as follows:
\begin{align}
\label{negativityDef}\mathcal{E}=&\sum_i\log \left(|\nu_i|+|1-\nu_i|\right)=\sum_i\log \left(\frac{|1+ \lambda_i|}{2}+\frac{|1- \lambda_i|}{2}\right).
\end{align}
This quantity is non-zero whenever there is a negative eigenvalue of the density matrix, $\nu_i<0$, which immediately means that there is also an eigenvalue large than $1$ since the eigenvalues come in pairs $(\nu,1-\nu)$. In terms of $\lambda$ this would mean an eigenvalue outside the range $[-1,1]$. In addition complex eigenvalues (either $\nu$ or $\lambda$ are complex) lead to non-zero negativity. We will find here that in the situation we consider eigenvalues may become complex but not negative.

\section{Orthogonal Polynomial Approach}
To determine the eigenvalues of the deformed correlation matrix $\bm f$, we must compute the determinant of the matrix 
$\mathcal{P}_{A\cup B}(\lambda+ {\bm f})$
The determinant is denoted by $D_{kln}$, \begin{align}
D_{kln}=\det\mathcal{P}_{A\cup B}(\lambda+ { \tilde f})\mathcal{P}_{A\cup B}\label{lambdadefdef}=\prod_{i=1}^{k+l} \left(\lambda + \lambda_i^{}\right),
\end{align}
the determinant being the characteristic polynomials.

We now decompose the matrix $\mathcal{P}_{A\cup B}(\lambda+\tilde f)\mathcal{P}_{A\cup B}$ in an upper-lower triangular decomposition as follows 
\begin{align}
&\mathcal{P}_{A\cup B}(\lambda+\tilde f)\mathcal{P}_{A\cup B}B=C
\end{align}
$B$ as upper triangular with all entries being given by unity on the diagonal, and $C$ as lower triangular with arbitrary  elements on the diagonal. For the following we truncate the matrices such that they are assume to have size $k+l+2$ (the size of intervals $A$ and $B$ are $k+1$\ and $l+1,$ respectively) corresponding to the size of $A \cup B$, whereby by convention we choose the elements corresponding to $A$ to precede those corresponding to $B$.   

Denoting the $i$-th column of $B$  by $B^{(i)}$we see that it obeys the equations:
\begin{align}
&B^{(i)}_{i}=1\label{BNmonomial}\\
&B^{(i)}_{j}=0\mbox{ for } j>i\label{BNPolynomial}\\
&((\lambda+\tilde f)B^{(i)})_j=0\mbox{ for }j<i,\label{BNResult}
\end{align}
One may convince oneself  that these equations do not depend on the size of the intervals $A$\ and $B$,  the conclusion being that the matrix elements of the $B$ and thus  $C$ do not depend on the size of the intervals (so long as they are not truncated out), it is only the question of which elements are defined that depend on the size of these intervals. Now since the determinant that we are seeking is given by the following equality 
\begin{align}
\det\left[{\mathcal{P}_{A\cup B}(\lambda+\tilde f)} \mathcal{P}_{A\cup B}\right]=\prod^{k+l+2}_{i=1}C_{ii},\label{detIsProdC} \end{align}
one may conclude that it is only the diagonal of the matrix $C$\ that must be extracted in order to find the characteristic polynomial of ${\mathcal{P}_{A\cup B}(\lambda+\tilde f)} \mathcal{P}_{A\cup B}.$ We designate $\chi$ as \begin{align}\label{chiIsC}\chi_{kln2+}\equiv C_{k+l+2,k+l+2}\end{align} making the dependence on $k,$ $l$ and $n$ explicitly. If we replace the order in which the elements in the matrices are placed, making those corresponding to $A$ follow those that correspond to $B$, then we have $\chi_{kln1+}\equiv C_{k+l+2,k+l+2}$, and if we invert the sense in which elements are placed within each interval such that instead of having sites to the left preceding those that are to the right, we have the ones to the left follow those to the right then we we obtain $\chi_{kln1-}$ and $\chi_{kln2-}$ instead being given by the last diagonal element of $C$. Rather than making these definitions more explicit here, we defer to Eqs. (\ref{Pchidef},\ref{chidef}) below for a more compact definition.

Combining Eq. (\ref{chiIsC}) and Eq. (\ref{detIsProdC}) allows us to write the first of the equations below\begin{align}
&  \quad\frac{D_{k,l,n}}{D_{k,l-1,n}}=\chi_{kln2+,\quad } \frac{D_{k,l,n}}{D_{k-1,l,n}}=\chi_{kln1+,}\label{DetsAndChis}
\end{align}
while the second equation gives $\chi_{kln1+}$ similarly to  $\chi_{kln2+}.$ The former is defined by an equation similar to Eq. (\ref{chiIsC}) but after  interchange of the order of the intervals in the matrix. 

It is useful to define the procedure of upper-lower triangular decomposition in terms of generating functions, which leads to the concept of orthogonal polynomials. Indeed, define:
\begin{align}
& {\bm \psi}^{kln}_{2 +}(z)=\begin{pmatrix}\sum_{i=0}^{k} z^iB_{k+l+2,i}\\
\sum_{i=0}^{l} z^iB_{_{k+l+2,i},i+k+1}\\
\end{pmatrix}.
\end{align}

More generally we define for any  $\sigma$ and $\omega$ where
$\omega\in\{+,-\}$ and $\sigma\in\{1,2\},$ the following vector:
\begin{align}
{\bm \psi}^{kln}_{\sigma \omega}=\left(\begin{array}{c}
\psi^{kln}_{1 \sigma \omega}\\
 \psi^{kln}_{2\sigma\omega}\end{array}\right).
\end{align}
The elements of the vector satisfy:
\begin{align}
\label{Pchidef} \psi^{kln}_{\sigma \omega}(z) \mbox{ }\begin{array}{lr}
\mbox{ $\psi^{kln}_{\sigma \sigma \omega}(z)$ is monomial of degree }m_\sigma  & \mbox{if } \omega=+1 \\ \\
\mbox{satisfies } \psi^{kln}_{\sigma \sigma \omega}(0)=\chi^{-1}_{kln\sigma -}& \mbox{if } \omega=-1\\
\end{array},
\end{align}
where
 \begin{align}
&  m_1=k,\,m_2=l. \label{ms}
\end{align}
The functions $\psi^{kln}_{\alpha\sigma\omega}(z)$ for $\alpha\neq\sigma$ are  polynomials of degree $m_\alpha$.
We demand the following property of $\bm \psi$ makes it into a vector orthogonal polynomial:\begin{align}
&\bm{e}_{\sigma'}\int  z^{-j }{\bm f}(z;\lambda) {\bm \psi}^{kln}_{\sigma \omega}(z) \frac{d\theta}{2\pi} =\chi^{\frac{\omega+1}{2} }_{kln\sigma\omega}\delta_{\sigma,\sigma'}\delta_{j,\frac{\omega+1}{2} m_\sigma}\label{chidef}
\end{align}
 here $\frac{\omega+1}{2}$ is an exponent not an index, and\ $\bm e_\sigma$ is the unit vector in direction $\sigma,$ namely the $\sigma'$ element of $\bm e_\sigma$, denoted by $e_{\sigma\sigma'}$ is given by:
\begin{align}
e_{\sigma \sigma'}=\delta_{\sigma\sigma'}.
\end{align}

For $\sigma=2$ and $\omega=+,$ the first line of Eq. (\ref{Pchidef}), is just the Fourier transform of  Eqs.(\ref{BNPolynomial}-\ref{BNResult}),
 while Eq. (\ref{chidef})  is the Fourier transform of Eq. (\ref{BNmonomial}) ,
given the identification of $C_{k+l+2,k+l+2}$ as $\chi_{kln2+}, $ Eq. (\ref{chiIsC}).  A similar conclusion may be made concerning the identification of $\chi_{kln1+}$ with the second ratio of determinants in Eq. (\ref{DetsAndChis}).

\section{The Riemann-Hilbert Problem}
The conditions on the orthogonal polynomials can be encoded by a Riemann-Hilbert problem that then may be solved (following first Refs. \cite{Deift:Its:Krasovksy:Toeplitz:Hankel,Deift:Its:Krasovsky,Deift:Its:Krasovsky:Toeplitz} and then more specifically to the problem at hand the prequel to this work, namely Ref. \cite{Bettelheim:Banerjee:Plenio}). To do so, we define a matrix $T$ as follows:\begin{align}
&T(z)\label{Tdef}=\\&=\nonumber
\left(\begin{array}{cccc}
 \psi_{11+}^{k(l-1)n} & \psi_{21+}^{k(l-1)n} & \oint\frac{ {-\bm e}_2 {\bm f}^{}(\xi;\lambda) {\bm \psi^{k(l-1)n}_{1+}}(\xi)}{(z-\xi)2\pi\imath\xi^l}  & \oint\frac{ {-\bm e}_1 {\bm f}^{}(\xi;\lambda) {\bm \psi^{k(l-1)n}_{1+}}(\xi)}{(z-\xi)2\pi\imath\xi^k} \\
\psi_{12+}^{(k-1)ln} &  \psi_{22+}^{(k-1)ln} &   \oint\frac{ {-\bm e}_2 {\bm f}^{}(\xi;\lambda) {\bm \psi^{(k-1)ln}_{2+}}(\xi)}{(z-\xi)2\pi\imath\xi^l}  &  \oint\frac{ {-\bm e}_1 {\bm f}^{}(\xi;\lambda) {\bm \psi^{(k-1)ln}_{2+}}(\xi)}{(z-\xi)2\pi\imath\xi^k} \\
-\psi_{12-}^{(k-1)(l-1)n} & -\psi_{22-}^{(k-1)(l-1)n} &  \oint\frac{ {\bm e}_2 {\bm f}^{}(\xi;\lambda){\bm\psi}^{(k-1)(l-1)n}_{2-}(\xi)}{(z-\xi)2\pi\imath\xi^l}  &  \oint\frac{ {\bm e}_1 {\bm f}^{}(\xi;\lambda){\bm\psi}^{(k-1)(l-1)n}_{2-}(\xi)}{(z-\xi)2\pi\imath\xi^k}
 \\
-\psi_{11-}^{(k-1)(l-1)n} & -\psi_{21-}^{(k-1)(l-1)n} &  \oint\frac{ {\bm e}_2 {\bm f}^{}(\xi;\lambda){\bm\psi}_{1-}^{(k-1)(l-1)n}(\xi)}{(z-\xi)2\pi\imath\xi^l}  &  \oint\frac{ {\bm e}_1 {\bm f}^{}(\xi;\lambda){\bm\psi}^{(k-1)(l-1)n}_{1-}(\xi)}{(z-\xi)2\pi\imath\xi^k}
\end{array} \right).
\end{align}
 It is then useful to define another matrix $Y$ as follows:
\begin{align}
Y_+(z)=T(z),\quad Y_-(z)=T(z)\begin{pmatrix}z^{-k} &  &  &  \\
 & z^{-l} &  &  \\
 &  & z^l &  \\
 &  &  & z^k \\
\end{pmatrix}
\end{align}
where $Y_\pm(z)$ are defined for $|z|<1$ and $|z|>1$ respectively. The jump condition on $Y$ is given by:

\begin{align}\label{YHilbert}Y_+(z)=Y_-(z) V
\end{align}
where the jump matrix, $V$,  is given by:\begin{align}
V=&\left(\begin{array}{cccc}
z^k&0& \imath fz^{-m} &  \lambda+f \\
0&z^l& \lambda- f& \imath  f  z^{m} \\
0&0& z^{-l}&0 \\
0&0& 0 &z^{-k}
\end{array} \right).\label{FullJump}
\end{align}
Together with the condition $Y_+(z)\to \mathds{1}+O(1/z),  $ Eqs. (\ref{YHilbert},\ref{FullJump}) both define a Riemann-Hilbert problem and fully encodes (or in, other word, is completely equivalent to) the problem of orthogonal polynomials given in Eqs. (\ref{Pchidef},\ref{chidef}).

\subsection{Solution in the Outer Region}

We first fix some notations.  We assume a jump discontinuity of $f(z)$ at $z^{(j)}$ :
\begin{align}
f(z^{(j)} e^{\pm\imath 0^+})=f^{(j)}_{o/i.}
\end{align}
Namely $f^{(j)}_{o/i}$ are the limiting values of $f$ as one approaches $z^{(j)}$ from the right or the left.
Each jump discontinuity is actually a Fermi point. Indeed, $f(z)$ is the occupation number of the femions (see Eq. (\ref{FermiFunction})). We shall then also write $e^{\pm\imath p_F}$ for $z^{(j)}$ respectively for $j=1,2$. 

The Riemann-Hilbert problem defined in Eq. (\ref{YHilbert}) may be solved easily in the region outside the Fermi points $z^{(j)}$. To do so we first define:
\begin{align}
&\label{rbeta}
 \beta^{(j)}=\frac{1}{2\pi\imath}\log\frac{\lambda+f^{(j)}_i}{\lambda+f^{(j)}_o},&\tilde\beta^{(j)}=\frac{1}{2\pi\imath}\log\frac{\lambda- f^{(j)}_i}{\lambda- f^{(j)}_o},
\end{align}
along with  the functions
\begin{align}
g_-(z)=\prod_j \left(\frac{z-z^{(j)}}{z}\right)^{-\beta^{(j)}}, \quad g_+(z)=\prod_j \left(z-z^{(j)}\right)^{\beta^{(j)}},
\end{align}
while $\tilde g_\pm$ are defined in a similar way by replacing $\beta^{(j)}$ with $\tilde \beta^{(j)}$. We take
\begin{align}
g_+g_-F_+F_-=\lambda +f, \quad \tilde g_+\tilde g_-\tilde F_+\tilde F_-=\lambda -f.
\end{align}
Now compute explicitly:
\begin{align}
 F_\pm(z)=g^{-1}_\pm (z)e^{\pm \oint \frac{\log\left(\lambda+f(z')\right)dz'}{2\pi\imath(z'-z) }}, \quad \tilde F_\pm(z)=\tilde g^{-1}_\pm (z)e^{\pm \oint \frac{\log\left(\lambda-f(z')\right)dz'}{2\pi\imath(z'-z) }}
\end{align}

These  definition above allow one to  write  a  solution to the Riemann-Hilbert problem in the entire complex plane excluding the Fermi points. This solution reads somewhat inexactly: 
\begin{align}
&Y^{(\rm out)}_{-}\label{out-}=\begin{pmatrix}\frac{1}{g_- F_-} & 0 & \frac{-\imath fz^{l-m}}{g_- F_-} & 0 \\
0 &\frac{1}{\tilde g_-\tilde F_-}    & 0 & 0 \\
0 & \frac{z^{-l}}{\tilde g_+\tilde F_+} & \tilde g_-\tilde F_- & 0 \\
-\frac{z^{-k}}{g_+F_+} & 0 & \frac{\imath f z^{l-m-k}}{g_+F_+} & g_-F_- \\
\end{pmatrix}
\\&Y^{(\rm out)}_{+}\label{out+}=\begin{pmatrix}
\frac{z^k}{g_-F_-} & 0 & 0 & g_+F_+ \\
0 & \frac{ z^l}{ \tilde g_-\tilde F_-} & \tilde g_+\tilde F_+ &\frac{\imath f z^m}{  \tilde g_-\tilde F_-}\\
0 & \frac{1}{ \tilde g_+\tilde F_+} & 0 & \frac{ \imath f z^{m-l}}{ \tilde g_+\tilde F_+} \\
-\frac{1}{g_+F_+} & 0 & 0 & 0 \\
\end{pmatrix}.  
\end{align}
I\ these expression $f$ is to be extended away from the unit circle in order to make sense of this solution. Here we encounter the inexactness of this representation, since $f$ typically may  only be analytically continued to a finite region around the unit circle excluding the Fermi points. Nevertheless, the solution away form the unit circle may be taken to be the expression obtained  by setting all non diagonal terms in $Y^{(\rm out)}_{-}$ to $0$ and all the terms away from the anti-diagonal in  $Y^{(\rm out)}_{+}$ to zero. This is justified by arguing that  those terms rapidly tend to zero away from the unit circle anyway. This is due to the fact this terms contain terms of the form $z^{r}$ where $r$\ is of order $N$ and are always of the right sign to vanish rapidly. This prescription may sound  rather inexact but may be made rigorous  using known methods\cite{Deift:Zhou:Steepest:Descent,Deift:Its:Krasovksy:Toeplitz:Hankel} applied to this case already in Ref. \cite{Bettelheim:Banerjee:Plenio}.
We forgo the more rigorous treatment as it was already discussed in Ref. \cite{Bettelheim:Banerjee:Plenio} and the interested reader may refer to that publication.

\subsection{Inner Riemann-Hilbert Problem} 

After offering a solution to the Riemann-Hilbert problem in the region that  ignores the Fermi points, it is necessary to find a solution in the regions surrounding each of the Fermi points, $z^{(j)}$. Only after combining the two regions together one may obtain a systematic large $N$ expansion of the orthogonal polynomials.

\begin{figure}[h!!!]
\begin{center}
\includegraphics[width=10cm]{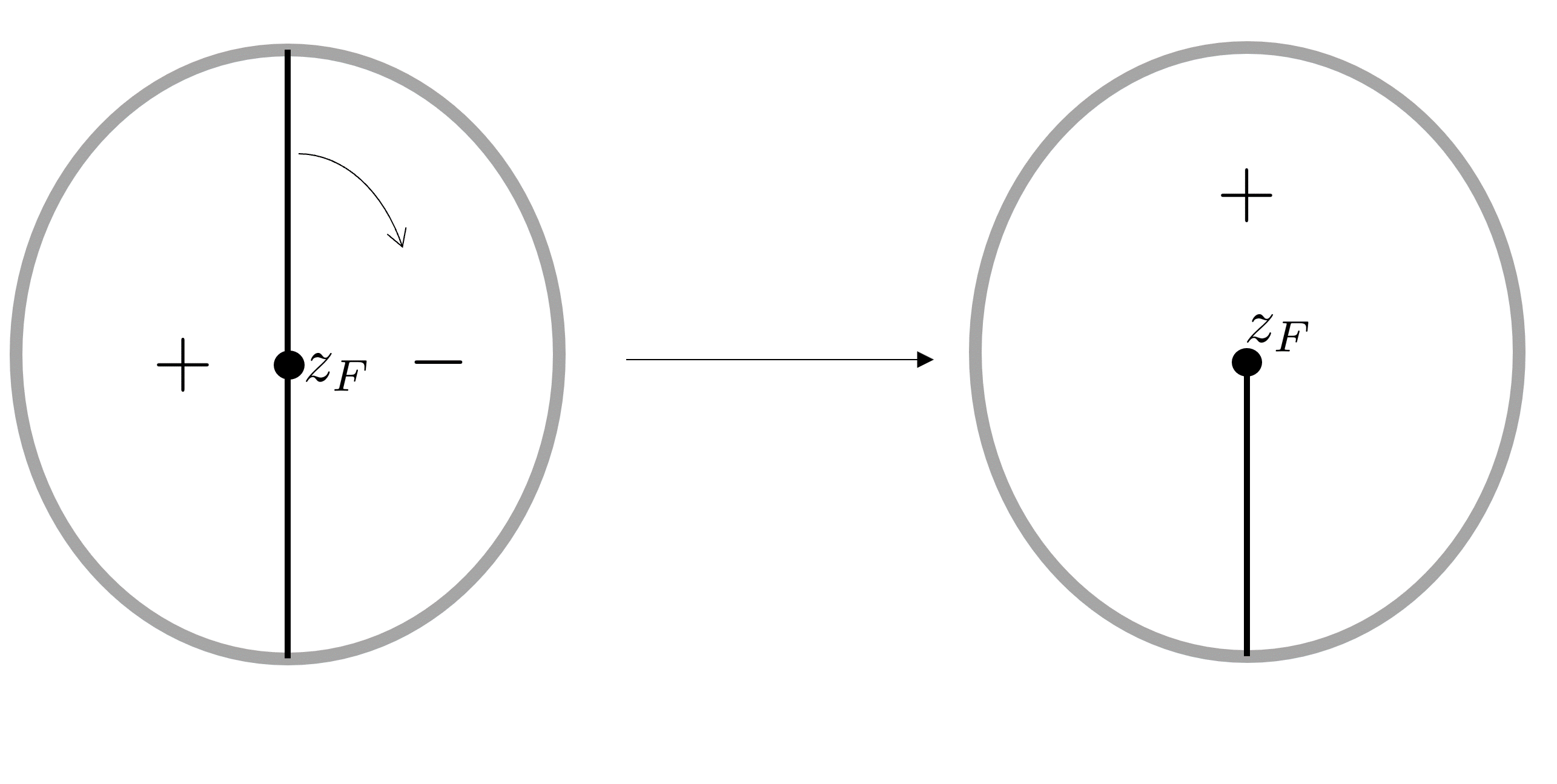}
\caption{
One may fold the line separating the $+$ and the $-$ regions (the inside and outside of the unit circle) such that it becomes a half line. The gray circles denotes an imaginary boundary between the inner region and the outer region, its exact location is arbitrary.  \label{InnerProcedure}} \end{center}
\end{figure}

In order to simplify the problem in the inner regions (the regions near the Fermi points), we make the following transformation:
\begin{align}
&y(z)=Y_+^{(\rm in)} (z)O_R^{}(z), \mbox{ for }|z|<1\\ & y(z)=Y^{(\rm in)}_-(z)O_L^{}(z), \mbox{ for }|z|>1 
\end{align} 
where the matrices $O_{R/L}^{}$ are given as follows: 
\begin{align}
&\label{OR}O^{}_R=\begin{pmatrix}\sqrt{2}z^{-(m+k)}  & \frac{z^{-(m-l+k)}}{\sqrt{2}} &  &  \\
\imath \sqrt{2} z^{-l}  & \frac{-\imath z^{-l}}{\sqrt{2}}  &  &  \\
 &  & \frac{\imath}{\sqrt{2}}  & \frac{-\imath}{\sqrt{2}}  \\
 &  & \frac{z^{-m}}{\sqrt{2}}  & \frac{z^{-m}}{\sqrt{2}}  \\
\end{pmatrix}  ,\\& \label{OL} O^{}_L=\begin{pmatrix}\sqrt{2} z^{-m} & \frac{z^{-m}}{\sqrt{2}}  &
\frac{\lambda z^{-m}}{\sqrt{2}}  & \frac{\lambda z^{-m}}{\sqrt{2}}  \\ \imath \sqrt{2}  & \frac{-\imath }{\sqrt{2}}  & \frac{\imath\lambda  }{\sqrt{2}}  & \frac{-\imath \lambda }{ \sqrt{2}}  \\
 &  & \frac{\imath z^{-l}}{\sqrt{2} }  & \frac{ -\imath z^{-l}}{\sqrt{2} }  \\
 &  & \frac{z^{-(m+k)}}{\sqrt{2}}  & \frac{z^{-(m+k)}}{\sqrt{2} }  \\
\end{pmatrix},\end{align}
and the superscript $(\mbox{in})$ denotes the object is defined for the inner region. The matrix $y$ can easily be seen to obey a Riemann-Hilbert problem:\begin{align}
y(e^{\imath x-0^+})=y(e^{\imath x+0^+})v\label{oldyeq},
\end{align}
 for $x$ real and where the jump matrix $v$ is related to the old jump matrix $V^{}$ through
\begin{align}V^{}O^{}_R=O^{}_L v .\end{align}
Explicitly, given the choice in Eqs. (\ref{OR},\ref{OL})  for the matrices $O^{}_R $ and $O^{}_L$, the jump matrix $v$ is given by:
\begin{align}
v\label{littlev}(z)=\begin{pmatrix}1 &  &  & f(z)  \\
 & 1 &  &  \\
 &  & 1 &  \\
 &  &  & 1 \\
\end{pmatrix}
\end{align}
$f(z)$ here is a function of $z$, but since a small circle is chosen around the Fermi point, where it has a jump discontinuity, it takes only two values which we have denoted by $f_i$ and $f_o$. If we look at the variable $\zeta=\log\frac{z}{z^{(j)}}$ This values are obtained on the positive and negative imaginary axis
as shown in Fig. \ref{InnerProcedure}. 

The  Riemann-Hilbert problem in this region consists in  finding a function which has the right asymptotics at large $\zeta$ to match the asymptotics of the outer region described in the previous section and that has the right jump discontinuity. The crux of the solving the latter, is to find a function that if follows its value starting in one of the regions, say the region $+$ as shown in Fig. \ref{InnerProcedure}, and then going around some curve that surrounds the origin, both jumps  according to the right jump matrix (at least twice, since we have a closed loop), but still manages to come back to its original value. Since this is the essential problem it is useful to mathematically turn the problem into just this one. This is done by folding the positive imaginary axis  clockwise onto the negative imaginary axis, until only the region $+$ remains and the region $-$ vanishes. The negative imaginary axis now entirely being responsible for the jump condition.  The procedure is shown in Fig. \ref{InnerProcedure}. The jump matrix obtained in this way is:
\begin{align}
v_\rightarrow =v_o\label{varrow}v^{-1}_i=\begin{pmatrix}1 &  &  & f_i-f_o  \\
 & 1 &  &  \\
 &  & 1 &  \\
 &  &  & 1 \\
\end{pmatrix}.   
\end{align}
and  the analytically extended $y(z)$ obeys the jump condition:
\begin{align}
y(e^{\zeta- 0^+})=y(e^{2\pi\imath}e^{\zeta+ 0^+})=y(e^{\zeta+ 0^+})v_\rightarrow,\label{newyeq}
\end{align}
for $\zeta$ on the negative imaginary axis (this equation to be compared to Eq. (\ref{oldyeq})). It is this problem that we solve in the following subsection.  

\subsection{Solution of the Inner Riemann-Hilbert Problem}
One may compute $y$ in the region $+$ from the form of $Y_{+}$ as follows (see Eq. (\ref{OR})):
\begin{align}
y=Y_{+}O_R^{}.\label{yDef}
\end{align}
 Now let us write for the $k$th row of $Y_{+}^{(k)}$  the following:
\begin{align}
\label{YIfromFGHJ}Y_{+}^{(k)}=\left(-\frac{(z^{(j)})^{m+k}}{\lambda+f_i} \frac{H(\zeta)}{\sqrt{2}},-\frac{\imath(z^{(j)})^l}{\lambda-f_i}\frac{F(\zeta)}{\sqrt{2}}  , \frac{\imath }{\sqrt{2}}G(\zeta),\frac{(z^{(j)})^{m}}{\sqrt{2}} J(\zeta)\right).
\end{align}
Crucially we do not allow the functions $H,$ $F,$ $G,$ or $J$ to have any dependce which is exponential in $N$ and  $\zeta$. Such terms will blow up in some region. Naively,  one could say that since $Y_+$ must only be well behaved in the $+$ region, terms that are vanishing in the $+$ region are acceptable. However, that even upon applying the jump matrix, those terms survive, and afflict $Y_-$ with divergencies, and are thus unacceptable.  We conclude that all these function must have at most power law divergence as $\zeta\to\infty$. 

Writing $\zeta=\log \frac{z}{z^{(j)}},$ one  obtains the following using Eqs.(\ref{yDef}, \ref{OR}):
\begin{align}&y^{(k)}_{+}\label{RequiredMonodromy}=\left(0,\frac{e^{-l\zeta }F-G}{2(\lambda-f_i)}+\frac{e^{-(m+k)\zeta }H-e^{-m\zeta }J}{2(\lambda+f_i)},\right.\\&\left.\nonumber,-\frac{G-e^{-m\zeta }J}{2}+\frac{e^{-l\zeta }F-G}{2(\lambda-f_i)}f_i,-\frac{G-e^{-m\zeta }J}{2}+\frac{e^{-l\zeta }F-G}{2(\lambda-f_i)} f_i\right)+\\
&\nonumber+\left(-\frac{G-e^{-l\zeta }F}{\lambda-f_i}+\frac{e^{-m\zeta }J-e^{-(m+k)\zeta }H}{\lambda+f_i},0,0, G\right)
\end{align}
The first row vector on the right hand side has to be analytic and the second has to obey the Riemann-Hilbert problem of Eq. (\ref{newyeq}, \ref{varrow}).
Indeed, the jump matrix, $v_\rightarrow$, specifies no jump discontinuity for vectors, the first element of which is zero, and as such, these vectors must be analytic.  
\subsection{Integrals}
We now present integrals which solve the Riemann-Hilbert problem for the inner region as presented above. 

Let $0\leq w\leq 3  $ and
\begin{align}
&t=\left(0,l,m,m+k\right)\\
&\gamma=\left( \tilde \beta-1+\delta_{0,w},-\tilde \beta-1+\delta_{1,w},\beta-1+\delta_{2,w},-\beta-1+\delta_{3,w}\right). 
\end{align}
 Note that $-t_i$ was already used in Fig. \ref{System} to denote the endpoints of the intervals. We consider the integrals:
\begin{align}
I_{i}^{[a,b]}(\zeta)=\int_a^b  e^{-t\zeta} \prod_{j=0}^3(t+t_i-t_j)^{-\gamma_j-1} dt
\end{align}

Let us consider the asymptotics of $I_{i}^{[a,b](\pm)}(\zeta),$ where $\pm$ signifies whether the integral is to be taken slightly above the cut of the integrand or slightly below namely: $I_{i}^{[0,\infty](\pm)}(\zeta)\equiv I_{i}^{[0,\infty]\pm\imath 0^+}(\zeta).$ Since it is possible to take the contour of integration to be the ray $[0,\infty] e^{\imath \theta}$, where $-\frac{\pi}{2}-\arg \zeta<\theta<\frac{\pi}{2}-\arg \zeta,$ then for $I_{i}^{[0,\infty](-)}(\zeta),$ one can choose   $\theta=0$ for $-\frac{\pi}{2}<\arg \zeta<\frac{\pi}{2}.$ For $\frac{\pi}{2}<\arg \zeta<\frac{3\pi}{2},$ one may choose $\theta=\frac{\pi}{2}-\arg \zeta.$ In the entire process of $\zeta$ acquiring a phase, a branch cut has not been crossed by the contour of integration (which is always in the lower half plane). The asymptote at $\zeta \to \infty$ is then:
\begin{align}
&I_{i}^{[0,\infty](-)}(\zeta)\label{InextoToLeading}=\Gamma(-\gamma_i)\zeta^{\gamma_i}\prod_{j\neq i}(t_i-t_j)^{-\gamma_j-1}\left(1-\sum_{j\neq i}\frac{\gamma_i(\gamma_j+1)}{(t_i-t_j)\zeta} +\dots\right)
\end{align}
for $-\frac{\pi}{2} <\arg \zeta < \frac{3\pi}{2}.$

We now consider the monodromy
\begin{align}
I^{[0,\infty](\pm)}_{i}(e^{2\imath \pi }\zeta )= I^{[0,\infty](\pm)}_{i}(\zeta )+ I^{[t_0-t_i,t_1-t_i](J)}_{i}(\zeta )+I^{[t_2-t_i,t_3-t_i](J)}_{i}(\zeta )
\end{align}
where the super $(J)$ denotes that the integral is rather of the  jump discontinuity of the relevant integrand $I^{[a,b](J)}_{i}=I^{[a,b](+)}_{i} -I^{[a,b](-)}_{i}$. In fact, by making use of the definition of $\beta$ and $\tilde{\beta}$, Eq. (\ref{rbeta}), this monodromy can be re-cast as:
\begin{align}
&\frac{e^{-t_i\zeta}I^{[0,\infty](-)}_{i}(e^{2\imath \pi }\zeta )-e^{-t_i\zeta}I^{[0,\infty](-)}_{i}(\zeta )}{f_i-f_o} =\\
&=\frac{e^{-t_3\zeta}I^{[0,\infty](-)}_{3}(\zeta )}{\lambda+f_i} -\frac{e^{-t_2\zeta}I^{[0,\infty](-)}_{2}(\zeta )}{\lambda+f_i}-\frac{e^{-t_1\zeta}I^{[0,\infty](-)}_{1}(\zeta )}{\lambda- f_i} +\frac{e^{-t_0\zeta}I^{[0,\infty](-)}_{0}(\zeta )}{\lambda- f_i}
\end{align}
Considering the form of the $k$th row of $Y_+$ given in Eq. (\ref{YIfromFGHJ}) and the required monodromy suggested in the text below Eq. (\ref{RequiredMonodromy})  we are lead to the following solution for $F$, $G$, $H$,$J$:
\begin{align}
\label{FGHJasIs}(F,G,H,J)=C_k\left(\frac{I^{[0,\infty](-)}_{1}(\zeta )}{(z^{(j)})^{l}} ,I^{[0,\infty](-)}_{0}(\zeta ),\frac{I^{[0,\infty](-)}_{3}(\zeta )}{(z^{(j)})^{m+k}} ,\frac{I^{[0,\infty](-)}_{2}(\zeta )}{(z^{(j)})^{m}} \right),
\end{align}
for some constant $C_k$ and for $w=s_k$ where
\begin{align}
s=\left(2,0,1,3\right).\label{ss}
\end{align} 
This is the solution to the Riemann-Hilbert problem in the inner region which we were seeking. We need now only to match asymptotics.  

\subsection{Matching Asymptotics}
We need the following asymptotics near $z^{(j)}$:
\begin{align} 
&\label{Youtexplicit}Y^{(\rm out)}_+(\zeta)=\begin{pmatrix}0 & 0 & 0 & e^{(j)}_1 \zeta^{\beta}  \\
0 & 0 & e^{(j)}_2\zeta^{\tilde \beta} & a \\
0 & e^{(j)}_3 \zeta^{-\tilde \beta} & 0 & a \\
e^{(j)}_4 \zeta^{-\beta} & 0 & 0 & a \\
\end{pmatrix}
\end{align}
where
\begin{align}
&\label{es}e^{(j)}=\left(d^{(j)},\tilde d^{(j)},-\frac{1}{\tilde d^{(j)}},-\frac{1}{d^{(j)}}\right)\\
&\label{ds}d^{(j)}=\prod_{j'\neq j}\left(z^{(j)}-z^{(j')}\right)^{\beta^{(j)}}F_+\left(z^{(j)}\right),  \quad  \tilde d^{(j)}=\prod_{j'\neq j}\left(z^{(j)}-z^{(j')}\right)^{\tilde \beta^{(j)}}\tilde F_+\left(z^{(j)}\right).
\end{align}
This is obtained from Eqs. (\ref{out-},\ref{out+}). 

If we write $z^{(1)}=e^{\imath p_F}$ and $z^{(2)}=e^{-\imath p_F}$ for $p_F>0$, then this can be written as:
\begin{align}
&d^{(j)}=\left(2e^{\imath \frac{\pi}{2}-\imath \pi\delta_{j,2}}\sin p_F\right)^{\beta^{(j)}}F_+\left(z^{(j)}\right),\\& \tilde d^{(j)}=\left(2e^{\imath \frac{\pi}{2}-\imath \pi\delta_{j,2}}\sin p_F\right)^{\tilde \beta^{(j)}}\tilde F_+\left(z^{(j)}\right).
\end{align}
\

Now comparing Eq. (\ref{YIfromFGHJ}, \ref{InextoToLeading}) gives us for $C^{(j)}$ in Eq. (\ref{FGHJasIs}):
\begin{align}
&C_k^{(j)}=c^{(j)}_k\frac{\prod_{i\neq s_k}(t_i-t_{s_k})^{\beta_i}}{\Gamma(-\beta_{s_k})}
\end{align}
where $s$ is given in Eq. (\ref{ss}) and
\begin{align}
&\label{cs}c^{(j)}=\sqrt{2}\left((z^{(j)})^{\ell_{02}}d^{(j)},-\imath \tilde d^{(j)},-\imath(z^{(j)})^{\ell_{01}}\frac{\tilde r}{\tilde d^{(j)}},(z^{(j)})^{\ell_{03}}\frac{r}{d^{(j)}}\right)
\end{align}
while $\arg (t_i-t_{s_k})^{\beta_i}=-\pi\beta_i$ for $t_i<t_{s_k}$.
Where
\begin{align}
\beta=(-\tilde\beta,\tilde\beta,\beta,-\beta)
\end{align}

All this allows us to write the asymptote of the whole matrix $Y_{+}$ explicitly at large $\zeta$ by making use of (\ref{InextoToLeading}). The small mismatch of the asymptotics between the inner and outer region is measured  by a matrix$R^{(j)}$:

\begin{align}
\label{KeyAsymptotics}
&Y^{(\rm in)}_+ Y^{(\rm out)-1}_+=\mathds{1}+\frac{R^{(j)}}{\zeta}.\end{align}
This matrix is crucial in finding corrections to the behvior of $Y$\ in the outer region. Indeed, we are interested in finding corrections of $Y_+(0)$ which are related simply to $T(0)$, which in turns hold valuable information about the orthogonal polynomials. In particular these elements will give us $\chi_{kln\sigma\omega}$.

The  elements of $R$ may be computed  using Eqs. (\ref{Youtexplicit},\ref{es},\ref{ds},\ref{cs}) on the one hand and Eqs. (\ref{YIfromFGHJ},\ref{FGHJasIs},\ref{InextoToLeading}), as these equations contain all the needed information about the asymptotics of $Y$ from the inner and outer regions. The result for the diagonal elements is:
\begin{align}
&R^{(j)}_{ii}\label{Rii}=\sum_{k\neq s_i}\frac{\beta^{(j)}_k\beta^{(j)}_{s_i}}{t_k-t_{s_i}}
\end{align} While non-diagonal elements have the form
\begin{align}
&R^{(k)}_{ij}=\frac{1}{t_{s_i}-t_{s_{j}}} \frac{c^{(k)}_i\Gamma(1-\beta^{(k)}_{s_{ j}})}{c^{(k)}_{ j}\Gamma(-\beta^{(k)}_{s_i})}\frac{\prod_{l\neq s_i}(t_l-t_{s_i})^{\beta^{(k)}_k}}{\prod_{l\neq s_j}(t_l-t_{s_j})^{\beta^{(k)}_l}}
\end{align}
 
From the mis-match between the outer solution, which ignores the inner regions, and the asymptotes in the inner region, one may find the correction to $Y^{\rm(out)}$. This procedure being fairly straightforward and covered already in Ref. \cite{Bettelheim:Banerjee:Plenio}, we write the final result for  $T_{ij}(0)$. This reads as follows:
\begin{align}
& T_{ij}(0)\label{T0usingR}= \left(\sum_k\delta_{i\bar j}+R^{(k)}_{i\bar{j}} \right)Y_{+\bar{j}j}^{(\rm out)}(0)
\end{align}
where $\bar j=5-j$.

\section{Computation of Mean Negativity Spectrum}
It turns out that it is quite easy from the formulas developed above to compute the density of eigenvalues $\lambda_i $ of the correlation matrix (see Eq. (\ref{lambdadefdef})). This amounts to setting probing the resolvent of the eigenvalues $\sum_i \frac{1}{\lambda+\lambda_i}$ or its pre-function $\log D_{kln}=\sum_i\log(\lambda+\lambda_i)$in the complex plane. The jump discontinuity of the resolvent represents the mean eigenvalue distribution. Nevertheless, the exact position of the individual eigenvalues may not be revealed by this computation if the order of limits is not taken properly. Namely if $\lambda$ is set to be a distance of order $1$ from the eigenvalue distribution, then the resolvent is computed in the $N\to\infty$ limit and only then the jump discontinuity of the resolvent is computed, then only the mean density of eigenvalues may be revealed, rather than the individual location of each of them. This presents a problem, since by using this order of limit a small imaginary value of the individual eigenvalues may be lost. In fact this turns out to be the case, and negativity may not be computed in this manner in our case. Nevertheless, we proceed with the computation, since it provides with an important test for the validity of our solution, as we may compare the mean density of eigenvalues obtained in the more naive order of limits with the density obtained below using a more intricate computation. The check is highly nontrivial since the computations turn out to be very different.

To compute the resolvent we compute the log determinant of the correlation matrix. We use Eq. (\ref{DetsAndChis}, \ref{chidef}) and  Eq. (\ref{Tdef}) to write:
\begin{align}
\chi_{kln1+}= \frac{D_{k,l,n}}{D_{k-1,l,n}}=\frac{\prod_i\lambda+\lambda_i^{k-1,l,n}}{\prod_i\lambda+\lambda_i^{k,l,n}}=T^{k,l+1,n}_{14}(0).
\end{align}
We may compute the very right hand side of this equation using Eqs. (\ref{Rii},\ref{T0usingR}) to give:
\begin{align} 
\log\chi_{kln1+}=\beta^2\left(\frac{1}{\ell_{21}}-\frac{1}{\ell_{20}}-\frac{1}{\ell_{23}}\right)
\end{align}
Similarly we have
\begin{align}
\log\chi_{kln2+}=\log \frac{D_{k,l,n}}{D_{k,l-1,n}}=T^{k,l+1,n}_{23}(0)=\beta^2\left(\frac{1}{\ell_{31}}-\frac{1}{\ell_{21}}-\frac{1}{\ell_{10}}\right)
\end{align}
This integrates into
\begin{align}
\log D_{kln}=\beta^2 \log\frac{\ell_{21}\ell_{30}}{\ell_{23}\ell_{20}\ell_{10}\ell_{31}}.
\end{align}
Computing the imaginary part of the jump discontinuity of the resolvent, (the resolvent being $\partial_\lambda \log D_{kln}=\sum_i \frac{1}{\lambda-\lambda_i}
$) one obtains the density of the eigenvalues:
\begin{align}
\rho(\lambda)=\log\frac{\ell_{21}\ell_{30}}{\ell_{23}\ell_{20}\ell_{10}\ell_{13}}\frac{1}{\pi}\Im[\beta (\imath0^-)\partial_\lambda \beta(\imath0^-)-\beta (\imath0^+)\partial_\lambda \beta(\imath0^+)]=\frac{\log\antifrac{\ell_{21}\ell_{30}}{\ell_{23}\ell_{20}\ell_{10}\ell_{13}}}{\pi^2(1-\lambda^2)}\label{EigenvalueDensity1}.
\end{align} 
This agrees with the result obtained below using very different methods in Eq. (\ref{EigenvalueDensity2}) up to $O(1)$ terms, here the result being exact to order $\log(N)$. Note also that in Eq. (\ref{EigenvalueDensity2}) the two intervals are assumed to be of the same size.  
 
\section{Computation of Fine Structure of Eigenvalues}

As mentioned above, it is not enough to compute the mean density of eigenvalues in order to compute logarithmic negativity, since a small imaginary value added to an eigenvalue will contribute to negativity but will not show up in the mean density of eigenvalues. One must thus approach the eigenvalue distribution a  microscopic distance away in order to sense the individual positions of eigenvalues. 

The problem which arises when one attempts to do this is that when $\lambda$ approaches the eigenvalue distribution,  then the real part of $\beta^{(1)}$  and $\tilde\beta^{(j)}$ approaches $\pm\frac12,$ namely $\Re\beta^{(j)}\to\pm \frac{1}{2},$ and $\Re\tilde \beta^{(j)}\to\mp \frac{1}{2}$, where the $\pm$ sign is chosen respectively with $j$. Thus we have that the difference in the real part of $|\Re (\beta^{(1)}-\beta^{(2)})|\to1$ and $|\Re (\tilde \beta^{(1)}-\tilde\beta^{(2)})|\to1,$ which is exactly the point at which the Riemann-Hilbert solution above fails, for the reasons that are given for example in Ref. \cite{Deift:Its:Krasovksy:Toeplitz:Hankel}. Following Ref. \cite{Deift:Its:Krasovsky}, we circumvent this problem by developing orthogonal polynomial identities, which map the problem of  finding the orthogonal polynomials for the case $|\Re (\beta^{(1)}-\beta^{(2)})|\to1$ and $|\Re (\tilde \beta^{(1)}-\tilde\beta^{(2)})|\to1$ to the problem of finding these polynomials for  $|\Re (\beta^{(1)}-\beta^{(2)})|\to0$ and $|\Re (\tilde \beta^{(1)}-\tilde\beta^{(2)})|\to0.$ This is done in the next subsection. 

Upon computing the eigenvalue distribution in this manner we compute logarithmic negativity by detecting the imaginary parts of the eigenvalues and applying that to the formula for logarithmic negativity, Eq. (\ref{negativityDef}).  
\subsection{Orthogonal Polynomials Idensitites Resovling  the Fine Structure} We now assume that $\bm f$ is characterized by $\Re \beta^{(j)}=-\frac12$ and $\Re \tilde\beta^{(j)}=\frac12.$ As such and according to Ref \cite{Deift:Its:Krasovksy:Toeplitz:Hankel} it may be dealt with using the Riemann-Hilbert approach as outlined below. However we are interested in the case by $\Re \beta^{(j)}=(-)^j\frac12$ and $\Re \tilde\beta^{(j)}=(-)^{j+1}\frac12.$   To achieve this we consider  $\hat{ \bm f}$  defined as follows:       
\begin{align}
\hat{\bm{f}}\equiv\begin{pmatrix}z & 0 \\
0 & z^{-1} \\
\end{pmatrix} \bm f ,
\end{align}
which effectively shifts say $\beta^{(1)}$ and $\tilde\beta^{(1)}$ by $\pm 1$ respectively. The orthogonal polynomials of $\hat{\bm f}$ are then computed not by associating a Riemann-Hilbert problem with them and then solving that Riemann-Hilbert problem, but rather by solving for the orthogonal polynomials associated with $\bm f$  and then using identities relating the orthogonal polynomials associatd with $\hat{\bm f}$ with those associated with $\bm f$ as shown below. 

We denote the orthogonal polynomials associated with $\hat{\bm f}$ as $\hat{\bm \psi}^{kln}_{\sigma \omega}$ and define the following:
\begin{align}
\hat{\bm \Psi}^{kln}_{\sigma\omega}=\begin{pmatrix}z & 0 \\
0 & 1 \\
\end{pmatrix} \hat{\bm \psi}^{kln}_{\sigma\omega}\label{PsiDef}
\end{align}
This leads to the following orthogonality condition:

\begin{align}
&\bm{e}_{\sigma'}\int  z^{-(j+\delta_{\sigma',2}) }\tilde{\bm f}^{}(z)\hat \Psi^{kln}_{i\sigma\omega} \frac{d\theta}{2\pi} =\chi^{\frac{\omega+1}{2} }_{kln\sigma\omega}\delta_{\sigma,\sigma'}\delta_{j,\frac{\omega+1}{2} m_\sigma}, \quad \label{PsiOrtho} 0\leq j\leq m_{\sigma'},
\end{align}
where \begin{align}\tilde {\bm f}=\begin{pmatrix}z & 0 \\
0 & 1 \\
\end{pmatrix} \bm f\begin{pmatrix}z & 0 \\
0 & 1 \\
\end{pmatrix} ^{-1}.\end{align} Actually $\tilde {\bm f}$ is dequal to $\bm f$ where $n$ is replaced with $n+1$, which will be useful below.  

From its definition, Eq. (\ref{PsiDef}), one concludes that $\hat{ \Psi}^{kln}_{ i\sigma \omega}$ has degree $k+1$ for $i=1$ and of degree $l$ for $i=2.$ Furthermore   $\hat{ \Psi}^{kln}_{ 1\sigma \omega}(0)=0.$ In addition plugging either ${\bm \psi}^{k,l,n+1},$   ${\bm \psi}^{k+1,l,n+1}_{ 1+}$  or    ${\bm \psi}^{k,l,n+1}_{ 2-}$ into the left hand side of Eq. (\ref{PsiOrtho})  for $\hat {\bm \Psi}^{kln}_{1+}$  (setting $\omega=+$ and $\sigma=1$) reproduces the right hand side for $0\leq j< m_{\sigma'}.$ One concludes that  $\hat {\bm \Psi}^{kln}_{1+}$   is a linear combination of  ${\bm \psi}^{k,l,n+1},$   ${\bm \psi}^{k+1,l,n+1}_{ 1+}$  and    ${\bm \psi}^{k,l,n+1}. $ In fact, the  appropriate linear combination is seen  to be given by 
\begin{align}
\label{psidoublehat}& \hat{\Psi}^{k,l,n-1}_{i1+}(z)=\frac{\det \begin{pmatrix}
\psi^{k,l,n}_{i1+}(z)&  \psi^{k+1,l,n}_{i1+}(z)& \psi^{k,l,n}_{i2-}(z) &  \\
\psi^{k,l,n}_{11+}(0)&  \psi^{k+1,l,n}_{11+}(0)& \psi^{k,l,n}_{12-}(0) & \\-T^{k,l+1,n}_{13}(0)&-T^{k+1,l+1,n}_{13}(0)&T^{k+1,l+1,n}_{33}(0)\\
\end{pmatrix}}{\det \begin{pmatrix}
\psi^{k,l,n}_{11+}(0)  & \psi^{k,l,n}_{12-}(0) & \\ T^{k,l+1,n}_{13}(0)&-T^{k+1,l+1,n}_{33}(0)\\
\end{pmatrix}}.
\end{align}
Indeed when one sets $z$ to zero, the first and second rows of the matrix in the numerator become identical and thus one obtains $\hat{\Psi}^{k,l,n-1}_{11+}(0)=0$,  which was mentioned above as a property of  $\hat{\Psi}^{k,l,n-1}_{1\sigma\omega}(z)$. Secondly, when one computes the right hand side of Eq. (\ref{PsiOrtho}) after substituting the ansatz Eq. (\ref{psidoublehat}) for $\sigma'=2$ and $j=l$ one may show that it does in fact vanish as required due to the  third row of the matrix in the numerator of  Eq. (\ref{psidoublehat}). The denominator of  Eq. (\ref{psidoublehat}) ensures of course  the  correct normalization. Namely,  that $\hat{\Psi}^{k,l,n-1}_{11+}(z)$ is monomial of degree $k+1$.

From Eq. (\ref{psidoublehat}) it is a matter of combining Eqs. (\ref{psidoublehat}, \ref{PsiDef}) to obtains:
\begin{align}
&\hat \chi_{k,l-1,n-1,1+}=-\chi_{k,l-1,n,1+}\frac{\det \begin{pmatrix}
   T^{k+1,l,n}_{11}(0)& T^{k+1,l,n}_{31}(0) & \\ T^{k+1,l,n}_{13}(0)&T^{k+1,l,1,n}_{33}(0)\\
\end{pmatrix}}{\det \begin{pmatrix}
T^{k,l,n}_{11}(0)  & T^{k+1,l,n}_{31}(0) & \\ T^{k,l,n}_{13}(0)&T^{k+1,l,n}_{33}(0)\\
\end{pmatrix}}.
\end{align}

\subsection{Computation of the Fine Structure}
\ Now let us define
\begin{align}
&\theta_{i}=t_i p_F-\arg\left[\Gamma\left(\frac{1}{2}-\imath \Im(\beta_i)\right)\right]+\sum_{k\neq i} \Im(\beta_k) \log (2\sin(p_F)\ell_{ik})\label{thetaidef}\\
&\theta_{ij}=\theta_i-\theta_j.\label{thetaDiff}
\end{align}
This allows us to write:
\begin{align}
&{\det \begin{pmatrix}
   T^{k+1,l,n}_{11}(0)& T^{k+1,l,n}_{31}(0) & \\ T^{k+1,l,n}_{13}(0)&T^{k+1,l,1,n}_{33}(0)\\
\end{pmatrix}}
=U\det \begin{pmatrix}
  \frac{ \cos(\theta_{23})}{\ell_{23}}& \frac{ \cos(\theta_{13})}{\ell_{13}} & \\ \frac{ \cos(\theta_{20})}{\ell_{20}}&\frac{ \cos(\theta_{10})}{\ell_{10}}\\
\end{pmatrix}\end{align}
where $U$ does not have zeros as $\lambda$ is varied and is thus irrelevant for the purpose of finding the eigenvalues.
\begin{align}
&4\det \begin{pmatrix}
  \frac{ \cos(\theta_{23})}{\ell_{23}}& \frac{ \cos(\theta_{13})}{\ell_{13}} & \\ \frac{ \cos(\theta_{20})}{\ell_{20}}&\frac{ \cos(\theta_{10})}{\ell_{10}}\\
\end{pmatrix}=\frac{2\cos( \theta _{12}+ \theta _{30})}{\ell_{23}\ell_{10}}-\frac{2\cos( \theta _{12}- \theta _{30})}{\ell_{13}\ell_{20}}+\\&+e^{\imath2 \theta _{23}}\frac{\ell_{21}\ell_{30}}{\ell_{23}\ell_{10}\ell_{13}\ell_{20}}e^{\imath \theta _{12}+\imath \theta _{30}}+e^{-\imath2 \theta _{23}}\frac{\ell_{21}\ell_{30}}{\ell_{23}\ell_{10}\ell_{13}\ell_{20}}e^{-\imath \theta _{12}-\imath \theta _{30}}
\end{align}
Searching for zeros of the term on the left hand side, the quadratic equation in $e^{\imath\theta_{23}}$ can be solved  to yield:
\begin{align}
&e^{\imath(\theta _{23}+\theta_{10})}\label{PhaseResult}=\exp\imath \arccos\left[\frac{\ell_{32}\ell_{10}\cos( \theta _{12}- \theta _{30})}{\ell_{21}\ell_{30}}-\frac{\ell_{31}\ell_{20}\cos( \theta _{12}+ \theta _{30})}{\ell_{21}\ell_{30}}\right]
\end{align}
where 
\begin{align}
\ell_{ij}=|t_i-t_j|
\end{align}
The expression $e^{\imath(\theta _{23}+\theta_{10})}$ determined by Eq. (\ref{PhaseResult}) is a phase if the argument of the inverse cosine is smaller than one in absolute value. In that case all $\theta_i(\lambda),$ where $\theta_i$\ is a function of $\lambda$ due to the definition in Eq. (\ref{thetaidef}) which involves $\beta_k$, which itself is a function of $\lambda$. If $\theta_i(\lambda)$ is real then so is $\lambda$. Namely, what determines if we have real eigenvalues is whether the argument of the inverse cosine in Eq. (\ref{PhaseResult}) is smaller or larger than $1$ in absolute value.        

\subsection{Eigenvalue Distribution at the Decoupling Limit}
Eq. (\ref{PhaseResult}) is an equation for the position of the eigenvalues. We now compare the position of the eigenvalues when the two intervals are much further apart than their own typical size. In the remainder of this paper we shall assume that the state of the fermions is that of a zero temperature Fermi sea with a given fermi momentum $p_F$. This means that $f(e^{\imath p })=\pm 1$, where the plus sign is to be taken when $|p|<p_F$ and the minus sign otherwise. In this case $\tilde \beta=-\beta$.

In the limit where the two intervals are very far apart, Eq. (\ref{PhaseResult}) turns into:
\begin{align}
e^{\imath(\theta _{23}+\theta_{10})}=-e^{\pm\imath(  \theta _{12}+ \theta _{30}) },
\end{align}
which further simplifies into
\begin{align}
\theta _{23}=\pm\frac{\pi}{2}+2n\pi, \mbox{  or     } \theta _{10}=\pm\frac{\pi}{2}+2m\pi,
\end{align}
so that Eq. (\ref{PhaseResult}) becomes the following equations for the imaginary part of $\beta, $ which itself is a function of $\lambda$:
\begin{align}
&\pm\frac{\pi}{2}+2n\pi\label{decoupling1}=p_F\ell_{32}+2\Im(\beta)\log(2\sin(p_F)\ell_{23})+2\arg\Gamma\left(\frac{1}{2}-\imath \Im(\beta)\right),\\&
\pm\frac{\pi}{2}+2m\pi\label{decoupling2}=p_F\ell_{10}+2\Im(\beta)\log(2\sin(p_F)\ell_{10})+2\arg\Gamma\left(\frac{1}{2}-\imath \Im(\beta)\right).
\end{align} 
These equations allow us to ascertain that the position of the set of eigenvalues are just the union of the set of eigenvalues where only the first or the second interval exist. This is done by comparing our result, Eqs. (\ref{decoupling1}, \ref{decoupling2}) to that in Ref \cite{Deift:Its:Krasovsky}, where the same equation can be found for a single interval, the only difference being that in that reference only one equation exists for one interval.
The eigenvalues are then obviously also real since they are real for each interval separately. 

As the size of the intervals increase, eventually being of the same order as the distance between the intervals, the positions of the eigenvalues shift. They may form complex conjugate pairs after meeting. The eigenvalues that may coincide are those that correspond to the first and second solution of the inverse cosine in Eq. (\ref{PhaseResult}). These two solutions correspond, as was just ascertained, in the decoupling limit (of large distance between the intervals) to the eigenvalues corresponding to the first and the second interval, respectively. The two solutions collide  when the argument of the inverse cosine becomes unity.  

One can see that the coincidence of the two sets of solutions is rather non-universal and depends on the exact value of $p_F$ along a non-universal dependence on the size of the intervals. The non-universality comes from the first term in Eq. (\ref{thetaidef}) which can be seen to already cause spurious coincidence of the two sets of solutions in the decoupling limit, Eqs. (\ref{decoupling1},\ref{decoupling2}). Indeed when $p_F(\ell_{32}-\ell_{10})\mod 2\pi=0$ in the decoupling limit, we have a coincidence of the eigenvalues. This conditions is satisfied when the filling fraction of fermions, as measured by $\frac{p_F}{\pi} $, has the form $\frac{p}{q}$, where $p$ and $q$ relative primes, and $q$ being a divisor of $\ell_{32}-\ell_{10}$. When this condition is satisfied the eigenvalues are highly susceptible to any coupling between the intervals, and this may cause them to become complex. This is demonstrated on the case of half filling below.  

\subsection{Complex Eigenvalues for Half Filling }

We now study the case where $p_F=\frac{\pi}{2}$, which may be considered half-filling. To further simplify the equations, we choose intervals of the same length, $\ell_{10}=\ell_{23}$. This leads to simple expressions for the different angles $\theta_{ij}$ appearing in Eq. (\ref{PhaseResult}). Indeed, we have:
\begin{align}
&\theta_{12}=-\ell_{21}p_F, \quad \theta_{30}=\ell_{30}p_F.
\end{align}
The term in the square brackets in Eq. (\ref{PhaseResult}) can be written as:

\begin{align}
&\frac{\ell_{31}\ell_{20}\cos( \theta _{12}+ \theta _{30})}{\ell_{21}\ell_{30}}-\frac{\ell_{32}\ell_{10}\cos( \theta _{12}- \theta _{30})}{\ell_{21}\ell_{30}}=\\
&=\left(2\frac{\ell_{20}\ell_{31}}{\ell_{21}\ell_{30}}-1\right)\sin\left(\ell_{21}\frac{\pi}{2}\right)\sin\left(\ell_{30}\frac{\pi}{2}\right)+\cos\left(\ell_{21}\frac{\pi}{2}\right)\cos\left(\ell_{30}\frac{\pi}{2}\right)
\end{align}
One can show the following:\begin{align}
2\frac{\ell_{20}\ell_{31}}{\ell_{21}\ell_{30}}-1>1.
\end{align}
Thus if $\ell_{21}$ and $\ell_{30}$ are both odd then we have complex eigenvalues as the expression inside the square brackets in  Eq. (\ref{PhaseResult}) is larger than $1,$ if both are even then the eigenvalues are real and if one is odd and the other even then again the eigenvalues are real, the peculiarity of this latter situation is that Eq. (\ref{PhaseResult}) takes the simple form:
\begin{align}
\theta_{23}+\theta_{10}\mod \pi=0.
\end{align}
This leads to an eigenvalue distribution where the distance between consecutive eigenvalues changes smoothly, rather than having a staggered nature. The staggered nature of consecutive distances in the generic case stems from the eigenvalues being basically a union of two sets of eigenvalues, each associated with the   different branch of the inverse cosine function in Eq. (\ref{PhaseResult}). In the decoupling limit these two sets are associated directly with each one of the intervals.

The example of half filling, $p_F=\frac{\pi}{2}$, and equal interval lengths, shows the non-universal nature of the imaginary part of the eigenvalues, as we see that the imaginary values depend on the parity of length of the individual intervals. In addition it is clear that moving away from half filling, further introduces non-universal features into the question of the complex eigenvalues, further complicating the situation.

Nevertheless, we compute the negativity of the situation where it exists, namely  in the half-filling equally sized intervals case. We assume that the gap between the intervals, the length of which is  $\ell_{21},$ has an odd number of  sites and the distance of between the two outermost points of the union of the two intervals, $\ell_{30}$, is also odd. In this case we may use first $e^{\imath \arccos(x)}=x+\sqrt{x^2-1} $ and then use:
\begin{align}
\left(2\frac{\ell^2_{20}}{\ell_{21}\ell_{30}}-1\right)+\sqrt{\left(2\frac{\ell^2_{20}}{\ell_{21}\ell_{30}}-1\right)^2-1}=\frac{\ell_{30}}{\ell_{21}}.
\end{align}
to reduce Eq. (\ref{PhaseResult}) into:
\begin{align}
\theta _{23}+\theta_{10}\mod 2\pi=\mp\imath \log\frac{\ell_{30}}{\ell_{21}}.
\end{align}  
Then using Eq. (\ref{thetaidef}) we have the following equation for the location of the eigenvalues in terms of $\beta_I(\lambda)\equiv\frac{1}{2\pi}\log\frac{1-\lambda}{1+\lambda}$ 
\begin{align}
&\mp\imath \log\frac{\ell_{30}}{\ell_{21}} \mod 2\pi=4\arg\left[\Gamma\left(\frac{1}{2}+\imath \beta_I)\right)\right]+2 \beta_I\log\frac{\ell_{30}\ell_{21}}{4\ell^2_{20}\ell^2_{10}}\label{SimpleCaseKey}
\end{align}
A consequence of this equation is that the eigevalues, $\lambda_\pm$ are to leading order real with a  small imaginary value of.  To leading order in $\frac{1}{\log N}$ the following may be easily deduced:
\begin{align}
\frac{\Im(\lambda_\pm)}{1-\Re^2(\lambda_\pm)}=\frac{\pi\log\left[\frac{\ell_{30}}{\ell_{21}}\right]}{2\log\frac{\ell_{30}\ell_{21}}{4\ell^2_{20}\ell^2_{10}}}, \end{align} 
under the assumption that $4\beta_I\log(\beta_I)\ll \log(N).$ If we sum the square of the imaginary part of both roots, $\lambda_\pm,$ we get:
\begin{align}
\sum_\pm\Im^2(\lambda_\pm)=\frac{\pi^2(1-\lambda^2)^2\log^2\left[\frac{\ell_{30}}{\ell_{21}}\right]}{4\log^2\frac{\ell_{30}\ell_{21}}{4\ell^2_{20}\ell^2_{10}}},
\end{align}
where we have replaced $\Re(\lambda_\pm)$ with $\lambda$ as we may think of $\lambda_\pm$ as being a function of  $\lambda$ which is real, the real part of $\lambda_\pm$ being just $\lambda$ while the imaginary part of $\lambda_\pm$ is small and depends on the index being either $+$ or $-$.

Negativity may be computed as follows 
\begin{align}
\mathcal{E}_\nu\label{NegativtyOneQuadrople}=&\sum_\pm\log \left(\frac{|1+\lambda_\pm|}{2}+\frac{|1-\lambda_\pm|}{2}\right)=\sum_\pm\frac{1}{2}\frac{\Im^2(\lambda_\pm)}{1-\lambda^2}
\end{align}
The density of either species of eigenvalues (namely, those denoted by $\lambda_+$ and those denoted by $\lambda_-$) may also be deduced from Eq. (\ref{SimpleCaseKey}) to leading order by considering the required change in $\lambda$ on the right hand side of that equation to lead to a change in $2\pi. $ The result of this simple calculation gives \begin{align}
\label{EigenvalueDensity2}\rho(\lambda)=\frac{1}{\pi^2(1-\lambda^2)}\log\antifrac{\ell_{30}\ell_{21}}{4\ell^2_{20}\ell^2_{10}}.
 \end{align}
This agrees with the result of Eq.(\ref{EigenvalueDensity1}). Negativity is given by 
\begin{align}
\frac{1}{2}\int \frac{\sum_\pm\Im^2(\lambda_\pm)}{1-\lambda^2}\rho(\lambda)d\lambda=\frac{\pi^2\log^2\left[\frac{\ell_{30}}{\ell_{21}}\right]}{8\log^2\frac{\ell_{30}\ell_{21}}{4\ell^2_{20}\ell^2_{10}}}\int _{-1}^1d\lambda
\end{align}
so  that the final result is given in Eq.(\ref{FinalFinal}).

\section{Conclusion}
In this paper we have computed logarithmic negativity for free fermions, where the distance between the intervals is macroscopically large. The most general result is given in implicit form in Eq. (\ref{PhaseResult}) together with Eqs. (\ref{thetaidef},\ref{thetaDiff})  for the case of  zero temperature. The solution for that equation are the eigenvalues $\lambda_i$ of the correlation matrix, trivially related to the eigenvalues of the density matrix, $\nu_i$ in Eq. (\ref{nulambdanega}). Negative eigenvalues of $\nu_i$ (which means that $\lambda_i$ is outside the interval $[-1,1]$) or complex eigenvalues of $\nu_i$ (or equivalently of $\lambda_i$ lead to non-zero logarithmic negativity. 

The consequence of Eq. (\ref{PhaseResult}) is that the location of the eigenvalues including an imaginary value will appear depending on whether $p_F$ (the Fermi momentum, the other Fermi momentum being set to $-p_F$) may be written as $\pi \frac{p}{q}$ where $p$ and $q$ are relatively prime integers, and if on the value of the size of the intervals modulo $q$. This is non-universal behavior which lead to logarithmic negativity most probably not being a good entanglement measure for this case. 

Despite the non-universal nature of the result we compute it in a particular case. Namely that of $p_F=\pi$. The result for equal intervals and odd number of sites between the intervals. The result is given in Eq. (\ref{FinalFinal}). The result behaves as $\frac{1}{\log(N)}$, $N$ being the typical size of either intervals. 

We should note that the present method is unable to detect eigenvalues with exponentially small negative value in $N$ (or imaginary value for that matter). As such it is not possible to determine if logarithmic negativity takes an exponentially small value in $N$ as was suggested in \cite{Marcovitch:Retzker:Plenio:Reznik:Critical:Noncritical:Long:Range:Entanglement:} for bosonic systems.      

%

\end{document}